\documentclass[11pt]{article}             % Use with LaTeX2e
\begin{document}
\thispagestyle{plain}

\label{sh}

%\lfoot[\fancyplain{\ \\[1mm] \thepage}{\ \\[1mm]\thepage}]{\fancyplain{}{}}

\begin{center} %{\Large \bf
\begin{tabular}{c}
PURITY DEPENDENT UNCERTAINTY RELATION AND POSSIBLE ENHANCEMENT
\\[-1mm]
OF QUANTUM TUNNELING PHENOMENON
\end{tabular}
% }
\end{center}

\bigskip

\bigskip

\begin{center} {\bf
V. N. Chernega }
\end{center}

\medskip

\begin{center}
{\it
P.N.~Lebedev Physical Institute, Russian Academy of Sciences\\
        Leninskii Prospect, 53, Moscow 119991, Russia\\
\smallskip
E-mail:~~~vchernega@gmail.com\\}
\end{center}

\noindent{\bf Keywords:} uncertainty relations, purity parameter, tunneling, position-momentum covariance
%\begin{document}

%\maketitle

\begin{abstract}
The position-momentum uncertainty relations containing the dependence of their quantum bounds on state purity parameter $\mu$ are discussed in context of possibilities to influence on the potential barrier transparency by means of decoherence processes. The behavior of barrier transparency $D$ is shown to satisfy the condition $\mu^{-1}\ln D=const$. The particular case of thermal state with temperature $T$ where the purity parameter is a function of temperature is considered. For large temperature the condition for the barrier transparency is shown to be $T\ln D=const$.
\end{abstract}

\medskip

\section{Introduction}
%\pst
The position-momentum uncertainty relation by Heisenberg ~\cite{Heis27} provides the quantum bound determined by Planck constant for product of these conjugate variable variances. The uncertainty relation with the dependence of the bound for this product containing the covariance of the position and momentum was found by Schrodinger ~\cite{Sch30} and Robertson ~\cite{Rob29}. The bound increases for increasing covariance of the position and momentum. It means that both particle position and momentum can fluctuate stronger if their covariance is larger. The uncertainty relations with bound dependent on different state characteristics like, e.g. degree of nongaussianity were studied in ~\cite{CerfMandilara,CerfMandilara1}.
The deformed uncertainty relations were considered in ~\cite{Bastos}. The state-extended uncertainty relations were discovered in ~\cite{Trif1,TRrif2}. Problem of experimental check of the uncertainty relations was considered in \cite{Chernega,Chernega1}. Tomographic approach to study uncertainty relations was presented in \cite{FoundPhysRite}. The quantum fluctuations are responsible for such phenomenon as tunneling of a particle through a potential barrier ~\cite{LandauLifs,Gamov,Gamov1}. In this connection it was pointed out ~\cite{Kurmyshev80} that the covariance dependence of bound in Schrodinger-Robertson uncertainty relation could be interpreted as a formal increasing of Planck constant which becomes "effective Planck constant". The quantum tunneling phenomenon is completely determined by the Planck constant. In view of this in ~\cite{KlimTrudy} the possibilities to obtain strong quantum correlations of position and momentum which are available in correlated coherent states with large covariance of these conjugate variables were studied. It was shown that the parametric excitation of harmonic oscillator creates the correlated coherent states which appear due to modulation of the oscillator frequency. The correlated coherent states and the possibility to use the strong position-momentum correlations to enhance the tunneling through potential barrier were discussed in ~\cite{AdamenkoJETP1,AdamenkoJETP2}. In~\cite{183,DodJSO} it was shown that the bound for the product of the position and momentum variances increases for mixed states. The bound depends on state purity parameter. It could mean that there exists another possibility to influence the barrier transparency. Namely, decoherence processes decreasing the purity parameters could enhance the tunneling phenomenon. The aim of the work is to discuses decoherence process in the context of the purity dependent uncertainty relation ~\cite{183} and consider the possibility of influence on the tunneling phenomenon of the "effective Planck constant" which depends on the purity parameter and on the temperature in the case of the thermal states. The paper is organized as follows. In Sec. 2 we review the Schrodinger-Robertson uncertainty relations. In Sec. 3 we consider the temperature dependent bound in the uncertainty relations. In Sec. 4 we discuses the "effective Planck constant" and the influence of the large fluctuations of position and momentum on tunneling phenomenon.

\section{Uncertainty relations}
Let us derive standard uncertainty relations for position and momentum following to ~\cite{183,DodJSO}. Let us use obvious inequality
\begin{eqnarray}\label{eq.1.01}
\langle B^{\dag}B\rangle\geq 0.
\end{eqnarray}
Here we mean that for any operator $\hat{D}$ we have $\langle\hat{D}\rangle=$$Tr$$\hat{\rho}\hat{D}$ where $\hat{D}$ is arbitrary operator and $\hat{\rho}$ is quantum state density operator. The inequality (\ref{eq.1.01}) for one-dimensional matrix $\hat{D}$ which is simply a complex number $z$ means that the number $z^{\ast}z=|z|^2\geq 0$. If we average this inequality we have $\langle z^{\ast}z\rangle\geq 0$. Let us construct the operator $\hat{B}$ as a linear combinations of position $\hat{q}$ and momentum $\hat{p}$ operators
\begin{eqnarray}\label{eq.1.02}
\hat{B}=c_1\hat{q}+c_2\hat{p}-c_1\langle\hat{q}\rangle-c_2\langle\hat{p}\rangle.
\end{eqnarray}
Here $\langle\hat{q}\rangle$ and $\langle\hat{p}\rangle$ are the means of the position and momentum, respectively. The inequality (\ref{eq.1.01}) gives the inequality for the quadratic form with respect to complex numbers $c_1$ and $c_2$, namely
\begin{eqnarray}\label{eq.1.03}
\sum^2_{jk=1}c_j^{\ast}A_{jk}c_k\geq 0
\end{eqnarray}
with the matrix $A_{jk}$ of the quadratic form which reads
\begin{eqnarray}\label{eq.1.04}
A_{jk}=\left(
          \begin{array}{cccc}
            \sigma_{qq} & \sigma_{qp} \\
            \sigma_{qp} & \sigma_{pp} \\
          \end{array}
        \right)+ \frac{i\hbar}{2}\left(
          \begin{array}{cccc}
            0 & 1 \\
            -1 & 0 \\
          \end{array}
        \right).
\end{eqnarray}
Here the matrix elements of the matrix $A_{jk}$ are variances and covariance of position and momentum
\begin{eqnarray}
 &&\sigma_{qq}=\langle\hat{q}^2\rangle-\langle\hat{q}\rangle^2;
 \nonumber\\&&
 \sigma_{pp}=\langle\hat{p}^2\rangle-\langle\hat{p}\rangle^2;
 \nonumber\\&&
 \sigma_{qp}=\frac{1}{2}\langle\hat{q}\hat{p}+\hat{p}\hat{q}\rangle-\langle\hat{q}\rangle\langle\hat{p}\rangle
 \label{eq.1.05}.
\end{eqnarray}
The term with the imaginary unit $i$ is related to commutator $[\hat{q},\hat{p}]=i\hbar\hat{1}$. The quadratic form is nonnegative if and only if the matrix $A_{jk}$ has nonnegative eigenvalues. It means that all the diagonal elements of the matrix are nonnegative and the determinant of the matrix is also nonnegative. The condition for the determinant provides the inequality
\begin{eqnarray}\label{eq.1.06}
\sigma_{qq}\sigma_{pp}-\sigma_{qp}^2\geq \frac{1}{4} \hbar^2.
\end{eqnarray}
It is the Schrodinger-Robertson  ~\cite{Sch30,Rob29} uncertainty relation for position and momentum. This inequality can be presented in another equivalent form
\begin{eqnarray}\label{eq.1.07}
\sigma_{qq}\sigma_{pp}\geq \frac{\hbar^2}{4}\frac{1}{1-r^2}.
\end{eqnarray}
Here $r$ is correlation coefficient of the random position and momentum
\begin{eqnarray}\label{eq.1.08}
r=\frac{\sigma_{qp}}{\sqrt{\sigma_{qq}\sigma_{pp}}}.
\end{eqnarray}
In classical domain $\hbar=0$ and there is no bound $\hbar^2/4(1-r^2)$ for the product of the position and momentum variances. The Heissenberg uncertainty relation ~\cite{Heis27}
\begin{eqnarray}\label{eq.1.09}
\sigma_{qq}\sigma_{pp}\geq \frac{\hbar^2}{4}
\end{eqnarray}
follows from the inequality (\ref{eq.1.07}) if one takes $r=0$. The correlation coefficient $r$ characterises the degree of statistical dependence of random position and momentum. If the variables are statistically independent, i.e. $r=0$ there is no correlations. One can interpret the inequality (\ref{eq.1.07}) that the presence of correlations increases the bound or provides an "effective Planck constant" value
\begin{eqnarray}\label{eq.1.10}
\hbar_{eff}=\frac{\hbar}{\sqrt{1-r^2}}
\end{eqnarray}
which depends on correlation coefficient $r$. The inequality (\ref{eq.1.07}) can be rewritten in the form of Heisenberg inequality
\begin{eqnarray}\label{eq.1.11}
\sigma_{qq}\sigma_{pp}\geq \frac{\hbar_{eff}^2}{4}.
\end{eqnarray}
Such interpretation was discussed in ~\cite{183}. The correlation coefficient takes the values in the domain $-1<r<1$. For $|r|$ close to $1$ the effective  Planck constant satisfies the inequality
\begin{eqnarray}\label{eq.1.12}
\hbar_{eff}\gg\hbar.
\end{eqnarray}
The value of Planck constant $\hbar$ is responsible for purely quantum phenomenon such as quantum tunneling ~\cite{LandauLifs}. The above inequality (\ref{eq.1.12}) means that for larger constant $\hbar_{eff}$ the quantum tunneling effect has to be enhanced. This remark has been done in ~\cite{183} and developed in ~\cite{AdamenkoJETP1,AdamenkoJETP2}. In next section we discuss another mechanism of the possible enhancing the quantum effect of the tunneling using the uncertainty relation where the bound depends on the purity parameter of the quantum state.

\section{Purity dependent uncertainty relation}
The mixed state of the quantum system with density operator $\hat{\rho}$ has the characteristics which is called purity parameter $\mu$. This parameter reads
\begin{eqnarray}\label{eq.1.13}
Tr\hat{\rho}^2=\mu.
\end{eqnarray}
The pure state density operator $\hat{\rho}$ has the property $\hat{\rho}^2=\hat{\rho}$.
It means that for pure state $Tr\hat{\rho}^2=Tr\hat{\rho}=1$. One can use another parameter $s=1-\mu$ which is called linear entropy/ For pure state the linear entropy is equal to zero. The purity parameter satisfies the inequality $0<\mu\leq 1$. It was shown in ~\cite{183} that the uncertainty relation has the bound which depends on the purity parameter $\mu$, i.e.
\begin{eqnarray}\label{eq.1.14}
\sigma_{qq}\sigma_{pp}\geq  \frac{\hbar^2}{4(1-r^2) } \Phi^2(\mu).
\end{eqnarray}
Here the function $\Phi(\mu)_{(\mu=1)}=1$.
The function $\Phi(\mu)$ has different forms for different domains of the parameter $\mu$. The function $\Phi(\mu)$ reads ~\cite{183}
\begin{eqnarray}
 &&\Phi(\mu)=\Phi_1(\mu)=2-\sqrt{2\mu-1},
  \nonumber\\&&\mu_2\leq\mu\leq \mu_1,\quad \mu_1=1,\;\mu_2=\frac{5}{9};
 \nonumber\\&&
 \Phi(\mu)=\Phi_2(\mu)=3-\sqrt{8(\mu-\frac{2}{3})},
  \nonumber\\&& \mu_3\leq\mu\leq \mu_2, \quad \mu_3=\frac{7}{18}
 \label{eq.1.15}.
\end{eqnarray}
Such pieces of the function $\Phi(\mu)$ are available for infinite number of $\mu_k$.
Thus the function $\Phi(\mu)$ has  different explicit form for different intervals of purity parameter and this function has the approximate interpolating expression
\begin{eqnarray}\label{eq.1.16}
\Phi_{app}(\mu)=\frac{4+\sqrt{16+9\mu^2}}{9\mu}.
\end{eqnarray}
For $\mu\ll 1$ the function $\Phi(\mu)$ reads
\begin{eqnarray}\label{eq.1.17}
\Phi_{app}(\mu)\simeq\frac{8}{9\mu}.
\end{eqnarray}
The systems in thermal state with the temperature $T$ have the density operator of the form
\begin{eqnarray}\label{eq.1.18}
\hat{\rho}(T)=\frac{1}{Z(T)}e^{-\frac{\hat{H}}{T}}
\end{eqnarray}
where $\hat{H}$ is the Hamiltonian of the system.
The partition function $Z(T)$ is given by the trace
\begin{eqnarray}\label{eq.1.19}
Z(T)=Tr e^{-\frac{\hat{H}}{T}}.
\end{eqnarray}
The purity parameter $\mu(T)$ is expressed in terms of the partition function
\begin{eqnarray}\label{eq.1.20}
\mu(T)=\frac{Z(T/2)}{Z^2(T)}.
\end{eqnarray}
The uncertainty relation for thermal state has the following form
\begin{eqnarray}\label{eq.1.14a}
\sigma_{qq}\sigma_{pp}\geq  \frac{\hbar^2}{4(1-r^2) } \Phi^2(\frac{Z(T/2)}{Z^2(T)}).
\end{eqnarray}
For small and large temperatures one can evaluate the purity parameter $\mu(T)$ by using expression for the oscillator $(\hbar=m=w=1)$ partition function
\begin{eqnarray}\label{eq.1.21}
Z(T)=\frac{1}{2\sinh(\frac{1}{2T})}
\end{eqnarray}
Then for small temperature we have expression for the purity parameter
\begin{eqnarray}\label{eq.1.22}
\mu_o(T)=\tanh(\frac{1}{2T})
\end{eqnarray}
which for $(1/2T)\gg 1$ is close to $\mu(\infty)=1$.
For large temperature $T\gg1$ one has the purity parameter
\begin{eqnarray}\label{eq.1.23}
\tilde{\mu}(T)=\frac{1}{2T}
\end{eqnarray}
which is close to zero. Thus for large temperature and $r=0$ "the effective Planck constant" dependence on temperature is described by the ratio $\hbar_{eff}(T)/\hbar\sim T$. It is known that there is the possibility to make fluctuations of the position and momentum larger if the temperature is larger. It means that the volume of the domain in the space (or the domain in the momentum space) where a particle can be found increases when the temperature increases. This is another explanation how the high temperature influences the possibility to overcome the repulsive forces, e.g. repulsion of two charged particles when the large kinetic energy of the particles at high temperatures is sufficient to compensate the potential energy responsible for the particle repulsion. Using known formulae \cite{LandauLifs} for barrier transparency $D$ for high temperature the behaviour of the barrier transparency can be characterised by the satisfying the requirement $T\ln D=const$.

\section{Decoherence as the tool to enhance the tunneling process}
The mentioned in Introduction method to influence the tunneling by increasing the correlation of position and momentum was discussed in detail in ~\cite{AdamenkoJETP1,AdamenkoJETP2}. Discussed above method to influence the tunneling process by using increasing the temperature has clear physical reasons since the high temperature means the large velocity and kinetic energy of the particles which provides the possibility to overcome  the repulsion forces. On the other hand we interpret this influence by the temperature dependent quantum bound for the product of the position and momentum variances. We point out that this bound for variance product depends not only on the correlation coefficient $r$ of position and momentum and not only on the temperature $T$ in the case of thermal states but on purity parameter $\mu$ which is not reduced to the temperature in general situation. For small purity parameter one can characterise the behaviour of the barrier transparency $D$ by satisfying the condition $\mu^{-1}\ln D=const$. The obtaining the small purity is usually considered as result of a decoherence process. This process is considered as the process which provides difficulties in quantum technologies like quantum computing, etc. On the other hand one can apply the decoherence process which gives the state with very small purity parameter $\mu\ll 1$. Even if there is no thermal equilibrium with such a parameter as temperature the impurity of the state can be obtained by other processes different from heating. Now we formulate the main result of our work. We interpreted the generalized  Heisenberg and Schrodinger-Robertson uncertainty relations obtained in ~\cite{183} where the bound of position and momentum variances product was shown to depend on "effective Planck constant" as possible resource to enhance the quantum phenomenon of tunneling by using decoherence process. For thermal state the heating of the system is known as possible tool to increase the position and momentum fluctuations. This gives intuitively clear picture why the system has large portion of particles with high energy sufficient to overcome the potential barriers. We point out that another phenomenon different from influence of high temperature could be used to enhance the quantum tunneling. The decoherence processes giving the very mixed states from initially pure ones also could be used to influence the tunneling. Thus the decoherence phenomenon is another tool, in addition to heating and increasing the correlation of position and momentum in pure state, to enhance the tunneling. We want to emphasize that all these three methods are based on the position-momentum uncertainty relation which depends on Planck constant. The presence of the extra functions like $1/(1-r^2)$ and $\Phi^2(\mu)$ as factors in the bound can be interpreted as reason to get "effective Planck constant" which is much larger than the standard  one. One has to point out that though the uncertainty relation with the purity dependent bound was recently checked experimentally ~\cite{Bellini,Porzio} the influence on the tunneling phenomenon of the decoherence or position-momentum quantum correlation needs to be experimentally confirmed.

\section*{Acknowledgements}
This study was partially supported by the Russian Foundation for
Basic Research under Project 11-02-00456.


\begin{thebibliography}{99}



\bibitem{Heis27}W. Heisenberg, {\sl Z. Phys.},
%\textbf{43},
{\bf43}, 172 (1927).
\bibitem{Sch30}E. Schr\"odinger, {\sl Ber. Kgl. Akad. Wiss.}, Berlin,
%\textbf{24},
{\bf 24}, 296
(1930).
\bibitem{Rob29} H. P. Robertson, {\sl Phys. Rev. A},
%\textbf{35},
{\bf35}, N 5, 667
(1930).
\bibitem{CerfMandilara} A. Mandilara, E. Karpov, and N. J. Cerf, {\sl Phys. Rev. A},
%\textbf{79},
{\bf79}, 062302 (2009).
\bibitem{CerfMandilara1} A. Mandilara, and N. J. Cerf, {\sl Phys. Rev. A}, {\bf86},
%\textbf{86},
030102(R)(2012).
\bibitem{Bastos} C. Bastos, O. Bertolami, N. Costa Dias [et al.], {\sl Phys. Rev. D}, {\bf86},
%\textbf{86}
105030 (2012).
\bibitem{Trif1} D. A. Trifonov, {\sl J. Phys. A: Math. Gen.}, {\bf33},
%\textbf{33},
L299 (2000).
\bibitem{TRrif2} D. A. Trifonov,{\sl  Eur. Phys. J. B - Cond. Matter Complex Syst.}, {\bf29},
%\textbf{29}
349 (2002).
\bibitem{Chernega} V. N. Chernega, {\sl Phys. Scr.}, {\bf T147},
%\textbf{T147},
014006 (2012).
\bibitem{Chernega1} V. N. Chernega, and V. I. Man'ko,
%State-extended uncertainty relations and tomographic inequalities as quantum-system-state characteristics
{\ sl Int. J. Quantum Inf.}, at press (arXiv:1210.0464)
\bibitem{FoundPhysRite}M. A. Man'ko, and V. I. Man'ko, {\sl Found. Phys.}, {\bf41},
%\textbf{41}
330 (2011).
\bibitem{LandauLifs} L. D. Landau, E. M. Lifshitz, {\it Quantum mechanics, Vol. 3. Non-relativistic theory}, Pergamon (1991).
\bibitem{Gamov} G. A. Gamov, {\sl Zeitschrift fur Physik}, {\bf52},
%\textbf{52},
510 (1928).
\bibitem{Gamov1} G. A. Gamov, {\sl Zeitschrift fur Physik}, {\bf51},
%\textbf{51},
204 (1928).
\bibitem{Kurmyshev80} V. V. Dodonov,  E. A. Kurmyshev,  and V. I. Man'ko, {\sl
Phys. Lett. A}, {\bf79},
%\textbf{79},
150 (1980).
\bibitem{KlimTrudy} V. V. Dodonov, A. B. Klimov, and V. I. Man’ko, {\it Physical
effects in correlated quantum states. Squeezed and
Correlated States of Quantum Systems}, Proc. Lebedev
Physics Institute, vol. 205, ed. M. A. Markov, Commack:
Nova Science, %pp 61–107
(1993).
\bibitem{AdamenkoJETP1} V. I. Vysotskii, S. V. Adamenko, and M. V. Vysotskyy {\sl JETP}, {\bf115},
%\textbf{115}, No 4,
551 (2012).
\bibitem{AdamenkoJETP2} V. I. Vysotskii, M. V. Vysotskyy, and S. V. Adamenko {\sl JETP}, {\bf114},
%\textbf{114}, Issue 2,
243  (2012).
\bibitem{183} V. V. Dodonov, and V. I. Man'ko, {\it Invariants and Evolution of
Nonstationary Quantum Systems}, Proc. Lebedev Physics
Institute, vol. 183, ed. M.A.Markov, Commack: Nova Science, N. Y. (1989).
\bibitem{DodJSO} V. V. Dodonov, {\sl J. Opt. B: Quantum Semiclass. Opt.}, {\bf4},
%\textbf{4},
S98 (2002).
\bibitem{Bellini} M. Bellini, A. S. Coelho, S. N. Filippov, V. I. Man'ko, and A. Zavatta, {\sl Phys. Rev. A}, {\bf85},
%\textbf{85},
052129 (2012).
\bibitem{Porzio} V. I. Man'ko, G. Marmo, A. Porzio, S. Solimeno and F. Ventriglia, {\sl Phys. Scr.}, {\bf83},
%\textbf{83},
04500 (2011).
%\bibitem{Zavatta} Valentina~Parigi, Alessandro~Zavatta, Myungshik~Kim, Marco~Bellini {\sl Science}, {\bf 317}, 1890 (2007)
\end{thebibliography}
\end{document}